\documentclass[twoside,10pt]{article}
\usepackage{amsmath,amssymb}
\usepackage{xspace}
\usepackage{tam10}

\title{Various Facets of Spacetime Foam}
\shorttitle{Various Facets of Spacetime Foam}
\authors{Y.~Jack~Ng$^{1,2}$\email{yjng@physics.unc.edu}}
\shortauthors{Y.J.~Ng}
\affiliations{
$^1$ Department of Physics \& Astronomy, University of North Carolina,
\\ 
$~$~~~Chapel Hill, NC 27599-3255, USA
\\
$^2$ European Southern Observatory, Karl-Schwarzschild-Str. 2,
\\
$~$~~~85748 Garching, Germany
}

\abstract{Spacetime foam manifests itself in a variety of ways.
It has some attributes of a turbulent fluid.  It is the source of the 
holographic 
principle.  Cosmologically it may play a role in explaining 
why the energy density has the critical value, 
why dark energy/matter exists, and why the effective dynamical
cosmological constant has the value as observed.  Astrophysically
the physics of spacetime foam 
helps to elucidate why the critical acceleration in
modified Newtonian dynamics has the observed value; and it provides
a possible connection
between global physics and local galactic dynamics
involving the phenomenon of flat rotation curves of galaxies and the  
observed Tully-Fisher relation.
Spacetime foam physics also sheds light on nonlocal gravitational 
dynamics.   
}

\begin{document}
\maketitle

\section{Introduction}

Unity of physics dictates that various physical phenomena and the
principles underlying them are related to one another.  But some of
the concepts, phenomena and structures found in physics are more 
fundamental than others.  I believe spacetime foam (arising from 
quantum fluctuations of spacetime) 
belongs to the first (fundamental) category.  
In this talk I will show that spacetime foam has a multiplicity of 
sides and will argue \footnote{Some of the interpretation of the
physics given here may deviate from the orginal works I did with my
various collaborators.  I alone am responsible for such a
reinterpretation.} that it is the origin 
of some of the various phenomena we see around us. Spacetime foam
manifests itself in the holographic principle.  Its physics 
calls for a critical cosmic energy density and the existence of dark 
energy/matter.  At least partly it explains   
the observed critical 
galactic acceleration and it provides
an intriguing dark matter profile.
It has some attributes of a turbulent fluid.
And its physics may be related to the nonlocality of gravitational 
dynamics.
Each of these various facets of spacetime foam will be discussed in
a separate section below.

But first, let us examine how foamy spacetime is, or, in other words,
how large quantum fluctuations of
spacetime are.  This can be done by using the following two methods.

$\bullet$
The Wigner-Salecker experiment \cite{wig57,sal58,ng94,ng95}

To quantify the problem, let us consider the fluctuations of 
a distance $l$
between a clock and a mirror.  By sending a light signal from the clock to
the mirror and back to the clock in a timing experiment, we can determine 
$l$.   The clock's and the
mirror's positions jiggle according to Heisenberg's uncertainty principle,
resulting in an uncertainty $\delta l$.  From the jiggling of
the clock's position alone, the uncertainty principle yields
$(\delta l)^2 \geq \hbar l / mc$, where $m$ is the mass of the clock.
On the other hand, the clock must be large enough not to
collapse into a black hole; this requires
$\delta l \gtrsim 4Gm/c^2$, which combines with the requirement from 
quantum
mechanics to yield $(\delta l)^3 \gtrsim 4 l l_P^2$ (independent of the
mass $m$ of the clock), where $l_P = \sqrt{\hbar G/ c^3} \approx 
10^{-33}$cm is the Planck length.  
We conclude that {\it the fluctuation of a 
distance scales as its cube root} \cite{Karol}:
\begin{equation*}
\delta l \gtrsim l^{1/3} l_P^{2/3},
\end{equation*}
where we
have dropped multiplicative factors of order unity.  Henceforth we will
continue this practice of dropping such factors except in a couple of 
places.

$\bullet$
Mapping the geometry of spacetime\cite{llo04,gio04}

Let us consider mapping
out the geometry of spacetime for a spherical volume of radius $l$ over 
the
amount of time $T = 2l/c$ it takes light to cross the
volume.  One way to do this is to fill the space with clocks, exchanging
signals with the other clocks and measuring the signals' times of 
arrival.  This
process of mapping the geometry of spacetime is a kind of computation, in
which distances are gauged by transmitting and processing information.
The total number of operations, including the ticks of the clocks and
the measurements of signals, is bounded by the Margolus-Levitin
theorem \cite{mar98}
in quantum computation, which stipulates that the rate of operations for 
any
computer cannot exceed the amount of energy $E$ that is available for
computation divided by $\pi \hbar/2$.  This theorem, combined with the
bound on the total mass of the clocks to prevent black hole formation, 
implies that the total number of operations that can occur in this 
spacetime volume is no greater than
$2 (l/l_P)^2 / \pi$.  To maximize spatial resolution (i.e., to minimize
$\delta l$), 
each clock must tick
only once during the entire time period.  If we regard the operations
partitioning the spacetime volume into "cells", then on the average each
cell
occupies a spatial volume no less than $(4 \pi l^3 / 3) / (2 l^2 /\pi 
l_P^2)
\sim l l_P^2$, yielding an average separation between neighboring
cells
no less than $ \sim l^{1/3} l_P^{2/3}$.
\cite{hsu}  This spatial
separation is interpreted as the average minimum uncertainty in the
measurement of a distance $l$, that is, $\delta l \gtrsim l^{1/3}
l_P^{2/3}$, the same result as found above in the 
Wigner-Salecker gedanken experiment.  This result will be shown to be 
consistent with the holographic principle; hence the corresponding 
spacetime foam model is called the holographic model .

But there are many other models of spacetime foam \cite{FordHu}.
We can characterize them with a parameter $\alpha \sim 1$
according to $\delta l \sim l^{1 - \alpha}
l_P^{\alpha}$.  It is useful to introduce the following model
as a foil to the ($\alpha = 2/3$) holographic model. 
Instead of maximizing spatial resolution in the mapping of spacetime
geometry, let us consider spreading the spacetime cells uniformly
in both space and time.  In that case, each cell has
the size of $(l^2 l_P^2)^{1/4} =
l^{1/2} l_P^{1/2}$ both spatially and temporally so that each clock ticks
once in the time it takes to communicate with a neighboring clock.  Since
the dependence on $l^{1/2}$ has the hallmark of a random-walk 
fluctuation,
the (quantum foam) model corresponding to  $\delta l \gtrsim
(l l_P)^{1/2}$ is called the random-walk model \cite{AC}.
Compared to the holographic model, the random-walk model predicts a
coarser spatial resolution, i.e., a larger distance fluctuation,
in the mapping of spacetime geometry.\footnote{It
also yields a smaller bound on the information content in a spatial
region, viz., $(l/l_p)^2 / (l/l_P)^{1/2} = (l^2 / l_P^2)^{3/4} =
(l/l_P)^{3/2}$ bits.}  We will concentrate on the holographic model
--- the only correct model, in my opinion.  But 
occasionally we will consider the general
class of models parametrized by the different values of $\alpha$
(specifically only when we discuss the experimental/observational 
probing of spacetime foam).  Unless clarity demands otherwise, we
will put $c=1$ and $\hbar = 1$ henceforth.

\section{Spacetime foam and probing it with distant quasars/AGNs}

How can we test the spacetime foam models?
The Planck length $l_P \sim 10^{-33}$ cm is so short that we need an
astronomical (even cosmological) distance $l$ for its fluctuation $\delta
l$ to be detectable.  Thus let
us consider light (with wavelength $\lambda$)
from distant quasars or bright active galactic nuclei \cite{lie03,NCvD}.
Due to the quantum fluctuations of spacetime, the wavefront, while planar,
is itself ``foamy", having random fluctuations in phase \cite{NCvD} 
$\Delta \phi \sim 2 \pi \delta l / \lambda$. 
When $\Delta \phi \sim \pi$, the cumulative uncertainty in the
wave's phase will have effectively scrambled the wave front sufficiently
to prevent the observation of interferometric fringes.
Consider the case of PKS1413+135 \cite{per02}, 
an AGN for which the redshift is $z = 0.2467$.
With $l \approx 1.2$ Gpc and $\lambda = 1.6 \mu$m,
we \cite{NCvD} find $\Delta \phi \sim 10 \times 2 \pi$ and
$10^{-9} \times 2 \pi$ for the random-walk model and
the holographic model of spacetime foam respectively.
Thus the observation \cite{per02}
by the Hubble Space Telescope of an Airy ring for
this AGN {\it rules out the random-walk model} but fails to test the 
holographic model.

Furthermore we \cite{chr06} note that, due to
quantum foam-induced fluctuations in the phase,
the wave vector can acquire a cumulative random fluctuation in
direction with an angular spread of the order of
$\Delta \phi / 2 \pi$.  In
effect, spacetime foam creates a ``seeing disk'' whose angular diameter is
$\Delta \phi /(2 \pi)
\sim (l / \lambda)^{1 - \alpha} (l_P / \lambda)^{\alpha}$
for the model parametrized by $\alpha$. \footnote{This
is partly based on the intuition (or reasonable assumption)\cite{chr06} 
that 
spacetime foam fluctuations are isotropic such that the sizes of the 
wave-vector fluctuations perpendicular to and along the light of sight 
are comparable.  But we should keep in mind that this intuition, though
reasonable, could be wrong; after all, spatial isotropy is here
``spontaneously" broken with the detected light being from a particular
direction.}

For a telescope or
interferometer with baseline length $D$, this
means that dispersion (on the order of $\Delta \phi /2 \pi$ in the normal
to the wave front) will be recorded as a spread in the angular size of a
distant point source, causing 
{\it a reduction in the Strehl ratio, and/or the
fringe visibility when} $\Delta \phi / 2 \pi \sim \lambda / D$, i.e.,
\begin{equation*}
(l / \lambda)^{1 - \alpha} (l_P / \lambda)^{\alpha} \sim \lambda / D
\end{equation*}
{\it for a  diffraction limited telescope}.\footnote{
For example , for a quasar of 1 Gpc away, at an infrared wavelength of the
order of 2 microns, the holographic model of spacetime foam predicts a
phase fluctuation $\Delta
\phi \sim 2 \pi \times 10^{-9}$ radians.  On the other hand, an infrared
interferometer with $D
\sim 100$ meters has $\lambda / D \sim 5 \times 10^{-9}$.
Such an interferometer has the potential to test the holographic model 
with a bright enough quasar that distance away.}
Thus, in principle, for arbitrarily large distances spacetime foam sets a
lower limit on the observable angular size
of a source at a given wavelength $\lambda$.  Furthermore, the
disappearance
of ``point sources'' will be strongly wavelength dependent happening first
at short
wavelengths. Interferometer systems (like the Very Large Telescope
Interferometers
when it reaches its design performance) with multiple
baselines may have sufficient signal to noise to allow for the detection
of quantum foam fluctuations.
For a discussion of the constraints recent astrophysical
data put on spacetime foam models, see \cite{cfnp}.\footnote{See Ref. 
\cite{AC,GWinterf} for a discussion of using
gravitational-wave interferometers (like LIGO) or laser atom 
interferometers to detect spacetime foam.}

\section{Spacetime foam and turbulence}

John Wheeler \cite{wheeler}
was among the first to realize the connections between 
quantum gravity and the ubiquitous phenomenon of turbulence.
Due to quantum fluctuations, spacetime, when probed at very small scales, 
will appear very complicated --- something akin in complexity to a chaotic 
turbulent froth (which, as we all know, he dubbed spacetime foam, also known 
as quantum foam --- the subject matter of this talk.)  The connections 
between quantum gravity and turbulence are quite natural if we
recall the role of the (volume preserving) 
diffeomorphism symmetry in classical (unimodular) gravity and the volume 
preserving diffeomorphisms of classical fluid dynamics.  We may also 
recall that, in the case of irrotational fluids in three spatial 
dimensions, the equation for the fluctuations of the velocity potential 
can be written in a geometric form~\cite{unruh} with a 
metric having the canonical ADM form~\cite{unruh,abh}.  The upshot is 
that the velocity of the 
fluid $v^i$ plays the role of the shift vector in
Einsteinian gravity; a fluctuation of 
$v^i$ would imply a quantum fluctuation of the shift vector.

Furthermore, in fully developed turbulence in three spatial dimensions, 
the remarkable Kolmogorov scaling~\cite{kol} 
implies that $v$ scales with length scale $l$ as $\sim l^{1/3}$,
consistent with experimental observations.
On the other hand, according to the holographic model of spacetime foam, 
a distance $l$ fluctuates by an amount 
$\delta l \sim l^{1/3} l_P^{2/3}$.
If one defines a velocity as $v \sim \frac{\delta l}{t_c}$, where the 
natural characteristic time scale is $t_c \sim \frac{l_P}{c}$, then it 
follows that $v \sim c (l / l_P)^{1/3}$. \footnote{Here the speed 
of sound $c$ and the Planck length $l_P$ for an induced gravitational 
constant are effective quantities.}
Thus we have obtained a {\it Kolmogorov-like scaling 
in turbulence, i.e., the velocity scales as} 
\begin{equation*}
v \sim l^{1/3}.
\end{equation*}
Since the velocities play the role of the shifts, they describe how the 
metric fluctuates at the Planck scale.  The implication is that 
{\it at short distances, spacetime is a chaotic and 
stochastic fluid in a turbulent regime} with the Kolmogorov length 
$l$.~\cite{previous}

\section{Spacetime foam and the holographic principle}

In essence, the holographic principle\cite{wbhts,bekenstein,hawking}
says that although the world
around us appears to have three spatial dimensions, its contents can
actually be encoded on a two-dimensional surface, like a hologram.
In other words, the maximum entropy, i.e., the maximum number of degrees 
of freedom, of a region of space is
given by its surface area in Planck units.  In this section, we will
heuristically show that {\it the holographic principle has its origin in 
the quantum fluctuations of spacetime}.

Consider partitioning a spatial region measuring $l$ by $l$ by $l$
into many small cubes, with the small cubes being
as small as physical laws allow, so that we can associate one degree of
freedom with each small cube. \cite{found}
In other words, the number of
degrees of freedom that the region can hold is given by the number of
small cubes that can be put inside that region.

But how small can such cubes be?
A moment's thought tells us that each side of a small cube
cannot be smaller than the accuracy
$\delta l$ with which we can measure each side $l$ of the big cube.
Thus, the number of degrees of freedom (d.o.f.) in the region
(measuring $l$ by $l$ by $l$) is given by $l^3 / \delta l^3$, which,
since $\delta l \gtrsim l^{1/3} l_P^{2/3}$, is
\begin{equation*}
\# d.o.f. \lesssim (l / l_P)^2,
\end{equation*} 
as stipulated by the holographic principle.  Thus spacetime foam
manifests itself holographically.

\section{Spacetime foam and the critical cosmic energy density}

Assuming that there is unity of physics connecting the Planck scale to 
the
cosmic scale, we can now appply the holographic spacetime foam model to 
cosmology \cite{llo04,Arzano,plb} and henceforth we call that cosmology 
the holographic foam cosmology (HFC).  

Recall that the minimum $\delta l$ found for the
holographic model corresponds
to the case of maximum energy density $\rho = (3/ 8 \pi) (l l_P)^{-2}$
for a sphere of radius $l$ not
to collapse into a black hole.  Hence the holographic model, unlike
the other models, requires, for its
consistency, the energy density to have the "critical" value. \footnote{
By contrast, for instance, the corresponding energy density for
the random-walk model takes on a range of values:
$(l l_P)^{-2}\gtrsim \rho \gtrsim l^{-5/2} l_P^{-3/2}$.  (The upper bound
corresponds to the clocks ticking every $(l l_P)^{1/2}$ while the lower
bound corresponds to the clocks ticking only once during the entire time
$2l/c$.)}
Hence, {\it according to HFC, the cosmic energy density is 
given by}
\begin{equation*} 
\rho = (3 / 8 \pi) (R_H l_P)^{-2},
\end{equation*}
{\it where $R_H$ is the Hubble radius}.\footnote{Instead of the Hubble 
radius,
it has been suggested\cite{DP,GWCS} that one should perhaps use the 
Ricci's 
length.}
{\it This is the critical 
cosmic energy density as observed.} \footnote{For an alternative 
explanation of the observed value for $\rho$, see 
\cite{sorkin,unimod}.} \footnote {Note that $\rho$ depends
on the geometric mean of $R_H$, the largest length scale, and $l_P$, the
smallest length scale.  This indicates that there is an interplay or 
connection between ultraviolet and infrared dynamics in HFC and
in spacetime foam physics.}  Furthermore, since critical energy 
density is a hallmark of the inflationary universe scenario, HFC may
be consistent with (warm) inflation \cite{diego}.

\section{Spacetime foam and dark energy/cosmological constant}

In this section we will show that {\it HFC "postdicts" the existence of
dark energy and yields the correct magnitude of the effective
cosmological constant}. \cite{llo04,Arzano,plb}  
The argument goes as follows:
For the present cosmic era, the energy density is given by
$\rho \sim H_0^2/G \sim (R_H l_P)^{-2}$ 
(about $(10^{-4} eV)^4$), where $H_0$ is the present Hubble parameter.
Treating the whole universe as a computer, one can
apply the Margolus-Levitin theorem to conclude that the universe
computes at a rate $\nu$ up to $\rho R_H^3 \sim R_H l_P^{-2}$
($\sim 10^{106}$ op/sec), for a total of $(R_H/l_P)^2$ ($\sim10^{122}$) 
operations during its lifetime so far.
If all the information of this huge computer is stored in ordinary
matter, we can apply standard methods of statistical mechanics
\footnote{Recall that energy (which determines the number of operations) 
and entropy (which determines the number of bits) 
depend on the 4th and 3rd power of temperature respectively.}
to find that the total number $I$ of bits is $[(R_H/l_P)^2]^{3/4} =
(R_H/l_P)^{3/2}$ ($\sim 10^{92}$).
It follows that each bit flips once in the amount of time given by
$I/\nu \sim (R_H l_P)^{1/2}$ ($\sim 10^{-14}$ sec). However the average 
separation of neighboring bits is
$(R_H^3/I)^{1/3} \sim (R_H l_P)^{1/2}$ ($\sim 10^{-3}$
cm).  Hence, assuming only ordinary matter exists to store all the
information we are led to conclude that the time
to communicate with neighboring bits is equal to the time for each
bit to flip once.  It follows that the accuracy to which ordinary
matter maps out the geometry of spacetime corresponds exactly to the case 
of events spread out uniformly in space and time as for
the random-walk model of spacetime foam.

But, as argued in the introduction, the holographic model, not the
random-walk model, is the correct model of spacetime foam.
Furthermore, the sharp images of PKS1413+135
observed at the Hubble Space
Telescope have ruled out the latter model. From the 
theoretical as well as observational demise of the
random-walk model and the fact that ordinary matter only contains an 
amount of information dense enough to map out spacetime at a level
consistent with the random-walk model, one now infers that
spacetime is mapped to a finer spatial accuracy than that which
is possible with the use of ordinary matter. Therefore there must be 
another kind of substance with which spacetime
can be mapped to the observed accuracy, as given by the 
holographic model. The natural conclusion
is that {\it unconventional
(dark) energy/matter exists}!
Note that this argument does not make use of the evidence from recent 
cosmological (supernovae, cosmic microwave
background, and galaxy clusters) observations. 

Furthermore, the average energy carried by each constituent  
(particle/bit) of the unconventional energy/matter is \footnote{
Recall that $I \sim (R_H / l_P)^2$ for holographic foam cosmology.}
$\sim \rho R_H^3/I \sim R_H^{-1}$ ($\sim 10^{-31}$ eV).  Such
long-wavelength (hence ``non-local'') constituents of dark energy 
act as a {\it dynamical cosmological constant with the observed 
magnitude} \footnote{For an alternative
explanation of the observed magnitude of $\Lambda$, see 
\cite{sorkin,unimod}.}
\begin{equation*}
\Lambda \sim 3 H^2.
\end{equation*}
Thus HFC predicts an accelerating universe.
In order to have an earlier decelerating universe and to have a
cosmic transition from the decelerating expansion to a recent
accelerating expansion, one needs dark matter and probably also
an interaction between dark matter and dark energy \cite{pavon}.
\footnote {As argued in \cite{pavon}, an appropriate interaction
between the two components can even help to alleviate the 
cosmic coincidence problem.}

\section{Spacetime foam and critical galactic acceleration/MoND}

If holographic spacetime foam has provided the cosmos with an effective
cosmological constant, one wonders if it may also affect local galactic
dynamics.  In particular,
in view of Verlinde's recent proposal \cite{verlinde}
(see Appendix A) for the 
entropic \cite{bekenstein}, and thus
holographic \cite{wbhts}
reinterpretation of Newton's law, it is natural to ask:
can Newton's second law be modified by holographic spacetime foam 
effects?

We first have to recognize that we live in an 
accelerating universe (in accordance with HFC).
This suggests that we will need a generalization \cite{HMN} of 
Verlinde's proposal to de Sitter space 
with a positive cosmological constant which, according to HFC, is
related to the Hubble parameter $H$ by $\Lambda \sim 3 H^2.$
The Unruh-Hawking temperature \cite{Untemp}
as measured by a non-inertial observer 
with acceleration $a$ in the de Sitter space is given by
$ \sqrt{a^2+a_0^2}/(2 \pi k_B)$\, \cite{deser}, 
where $a_0=\sqrt{\Lambda/3} $ \,\cite{hawking}.
Consequently, we can define the net temperature measured by the 
non-inertial observer (relative to the inertial observer)
to be $\tilde{T} = [(a^2+a_0^2)^{1/2} - a_0]/(2 \pi k_B)$. 

We can now follow Verlinde's approach\cite{verlinde}.\footnote{
We replace the $T$ in Appendix A 
by $\tilde{T}$ for the Unruh temperature.}  Then the entropic 
force, acting on the test mass $m$ with acceleration $a$ in de Sitter 
space, is given by
$F_{entropic}=\tilde{T}\, \nabla_x S= m [(a^2+a_0^2)^{1/2}-a_0].$
For $a \gg a_0$, the entropic force is given by 
$F_{entropic}\approx ma$.  But
for $a \ll a_0$, we have
$F_{entropic}\approx ma^2/(2a_0)$; and so
the terminal velocity $v$ of the test mass $m$
should be determined from
\,$ m a^2/(2a_0) = m v^2/r$.\cite{HMN}
The observed flat
galactic rotation curves (i.e., at large $r$, $v$ is independent
of $r$) 
and the observed Tully-Fisher relation (the speed of stars
being correlated with the galaxies' brightness,
i.e., $v^4 \propto M$) 
\cite{dsmond} now require that
$ a \approx \left(\,4 \, a_N \,a_0^3 \, \right)^{\frac14}$,
where $a_N= G M /r^2$ is the magnitude of
the usual Newtonian acceleration.\footnote{One can check this
by carrying out a simple dimensional analysis and recalling that
there are two accelerations in the problem: viz, $a_N$ and $a_0$.
The factor of $4^{1/4}$ in $a$ is included for convenience only.}  
But that means
$ F_{entropic} \approx m a^2/(2a_0) \approx m \sqrt{a_N a_0}$ for the  
small acceleration $a \ll a_0$ regime.  Thus we are led to
the modified Newtonian dynamics, or MoND \cite{mond},
due to Milgrom, 
which stipulates that the
acceleration of a test mass $m$ due to the source $M$ is given by
$a= a_N$ and $\sqrt{a_N\, a_c}$ for $a \gg a_c$ and $a \ll a_c$  
respectively
\footnote{Our result is not surprising, since MoND has been designed
to give the observed flat rotation curves and the Tully-Fisher
relation in the first place.  Let us also note that actually
Milgrom suggested \cite{interpol} that the generalized Unruh 
temperature $\tilde{T}$ can give the correct behaviors 
of the interpolating function between the usual Newtonian
acceleration and his suggested MoNDian deformation for very small
accelerations.  He was right, but he could not offer any justification.}
--- provided we can identify $a_0$ as Milgrom's critical 
acceleration $a_c$.  
Milgrom has observed that 
$a_c$ is numerically related to the speed of light $c$ and
the Hubble scale $H$ as \footnote{To be more precise, $a_c \sim c H/(2
\pi)$.} $a_c \sim c H \sim 10^{-8} cm/s^2.$  
But $a_0 = (\Lambda /3)^{1/2}$, and $\Lambda 
\sim 3 H^2$ as argued in the last section for HFC, it follows that
$a_0$ is of the order of magnitude of 
\begin{equation*}
a_{critical} \sim \sqrt{\Lambda/3} \sim H.
\end{equation*}  
In other words, 
{\it we have 
successfully predicted the correct magnitude of the critical galactic
acceleration, and furthermore have found that global physics 
(in the form of a 
dynamical cosmological constant with its origin in spacetime foam) can 
affect local galactic motion!}

\section{Spacetime foam and cold dark matter with MoND scaling}

With only a single parameter
($a_c$), MoND can explain easily and rather successfully
(while the cold dark matter (CDM) paradigm cannot) the observed  
flat galactic rotation curves 
\footnote{Since the galactic dynamics is very complex, it is
not surprising that MoND {\it cannot explain
all} of the observed galactic velocity curves.}
\footnote{For other attempts
to explain the rotation curves of galaxies, see, e.g.,
\cite{Mann}; but typically they all make use of more than one
parameter.}
and the observed Tully-Fisher relation. 
But there are problems with MoND at the cluster and cosmological scales,   
where apparently CDM works much better \cite{dark}.
This inspires us \cite{HMN} to ask: Could there be some kind of dark 
matter that can behave like MoND at the galactic scale?  

Let us continue to follow Verlinde's holographic approach. 
Invoking the imaginary holographic screen of radius $r$, we can write
\footnote{We replace the $T$ and $M$ in Appendix A by 
$\tilde{T}$ and 
$\tilde{M}$ respectively.}
$2 \pi k_B \tilde{T} = \frac{G\,\tilde{M}}{r^2}$,
where $\tilde{M}$ represents the \emph{total} mass enclosed within the 
volume $V = 4 \pi r^3 / 3$.  
But, as we will show below, consistency with the discussion in the 
previous section (and with observational data) demands that
$\tilde{M} = M + M'$ where $M'$ is some unknown mass --- that 
is, dark matter. 
Thus, {\it we need the concept of dark matter for 
consistency.}  

First note that
it is natural to write the entropic force
$F_{entropic} = m [(a^2+a_0^2)^{1/2}-a_0]$ as
$F_{entropic} = m\,a_N [1 + 2 (a_0/a)^2]$ 
since the latter expression is arguably the simplest
interpolating formula \footnote{But it is not unique --
actually, it may be wrong for the $a \sim a_0$ regime.}
for $F_{entropic}$ that satisfies the two requirements: 
$a \approx (4 a_N a_0^3)^{1/4}$ in the small
acceleration $a \ll a_0$
regime, and $a = a_N$ in the $a \gg a_0$ regime.
But we can also write $F$ in another, yet equivalent, form:
$F_{entropic} = mG(M+M')/r^2$.
These two forms of $F$ illustrate the idea of CDM-MoND duality.\cite{HMN}
The first form can be interpreted to mean that there is no dark matter,
but that the law of gravity is modified, while the second form means
that there is dark matter (which, by construction, is consistent with 
MoND) but that the law of gravity is not modified.
The second form gives us this intriguing
dark matter profile: $M'=2 \,\left(\,\frac{a_0}{a}\,\right)^2 \, M$.
Dark matter of this kind can behave as if there is no dark matter 
but MoND.  Therefore, we call it ``MoNDian dark matter". \cite{HMN}
One can solve for $M'$ 
as a function of $r$ in the two acceleration regimes: 
$M' \approx 0$ for $a \gg a_0$, and (with $a_0 \sim \sqrt{\Lambda}$) 
\begin{equation*}
M' \sim (\sqrt{\Lambda}/G)^{1/2} M^{1/2} r
\end{equation*}
for $a \ll a_0$.
Intriguingly {\it the dark matter profile we have
obtained relates, at the galactic scale},
\footnote{One may wonder why MoND works 
at the galactic scale, but not at the cluster or cosmic scale.  One of 
reasons is that, for the larger scales, one has to use Einstein's equations  
with non-negligible contributions from the pressure and 
explicitly the cosmological constant, which have not been taken into 
account in the MoND scheme. \cite{HMN}}
{\it dark matter ($M'$), dark energy ($\Lambda$) and 
ordinary matter ($M$) to one another}.\footnote {
This requires all the three components to exist (an arguably welcome 
news to HFC) and it   
indicates possible interactions among them -- something, as observed
above, that 
we may need to alleviate the cosmic coincidence problem and to 
have a cosmic phase transition from a decelerating to an accelerating 
expansion at redshift $z \sim 1$. \cite{pavon}} 
As a side remark,
this dark matter profile can be used to recover the 
observed flat rotation curves and the Tully-Fisher relation.

\section{Spacetime foam and nonlocality}

According to the holographic principle, the
number of degrees of freedom in a region of space is bounded not by
the volume but by the surrounding surface.  This suggests that the
physical degrees of freedom are not independent but, considered
at the Planck scale, they must be infinitely correlated, with the result
that the spacetime location of an event may lose its invariant
significance.  {\it If we take the point of view that holography has its 
origin in spacetime foam} (as we have argued above), {\it then we can argue
that spacetime foam gives rise to nonlocality.}  This argument is also
supported by the following observation \cite{plb} that the long-wavelength
(hence ``non-local'') "particles" constituting dark energy in HFC
obey an exotic statistics which has attributes of nonlocality.

Consider a perfect gas of $N$ particles obeying Boltzmann statistics
at temperature $T$ in a volume $V$.  For the
problem at hand, as the lowest-order approximation, we can neglect the
contributions from matter and radiation to the cosmic 
energy density for the recent and present eras.  Then
the Friedmann equations for $\rho \sim H^2 /G$ can be solved by $H
\propto 1/a$ and $a \propto t$,
where $a(t)$ is the cosmic scale factor.
Thus let us take $V \sim R_H^3$, $T \sim R_H^{-1}$, and $N \sim
(R_H/ l_P)^2$. A standard calculation (for the relativistic case) yields 
the
partition function $Z_N = (N!)^{-1} (V / \lambda^3)^N$, where
$\lambda = (\pi)^{2/3} /T$, and the entropy
$S = N [ln (V / N \lambda^3) + 5/2]$.
The important point to note is that, since $V \sim \lambda^3$, the entropy
$S$ becomes nonsensically negative unless $ N \sim
1$ which is equally nonsensical because $N \sim (R_H/l_P)^2 \gg 1$.  The 
solution comes with the observation that
the $N$ inside the log term for $S$ somehow
must be absent.  Then $ S \sim N
\sim (R_H/l_P)^2$ without $N$ being small (of order 1) and S is 
non-negative
as physically required.  That is the case if the ``particles" are
distinguishable and nonidentical!  For in that case, the Gibbs $1/N!$ 
factor is absent from the partition function $Z_N$.
Now the only known consistent statistics in greater than two space
dimensions
without the Gibbs factor 
is infinite statistics (sometimes called
``quantum Boltzmann statistics") \cite{DHR,greenberg,CCL}.  (A
short description of infinite statistics is given in Appendix B.)
Thus we \cite{plb} have
shown that the ``particles" constituting dark energy obey infinite
statistics,
instead of the familiar Fermi or Bose statistics.
\footnote{Using the Matrix theory approach,
Jejjala, Kavic and Minic \cite{minic} have also argued 
that dark energy quanta obey infinite statistics.} 

But it is known that a theory of particles
obeying infinite statistics cannot be local \cite{fredenhagen,
greenberg}.  
The expression for the number operator 
\begin{equation*}
n_i = a_i^{\dagger} a_i + \sum_k a_k^{\dagger} a_i^{\dagger} a_i a_k   
+ \sum_l \sum_k a_l^{\dagger} a_k^{\dagger} a_i^{\dagger} a_i a_k a_l +
...,
\end{equation*}
is both {\it nonlocal} and nonpolynomial in the field operators,
and so is the Hamiltonian.
Altogether, the indication is that nonlocality is yet another facet of 
spacetime foam.\footnote{
An interesting question presents itself:  
Though the nonlocality in holography is probably related to the 
nonlocality in theories of infinite statistics, how exactly are
they related?} \footnote {The nonlocal nature of the dynamics of
gravitation has been pointed out in other contexts before, see, e.g., 
\cite{SG}.}

\section{Discussion}

In the above sections, we have discussed several facets of 
spacetime foam.  In this section we will mention one {\it non-}facet
of spacetime foam.

Motivated by the
interesting detection of a minimal spread in the arrival times of high
energy
photons from distant GRB reported by Abdo et al,\cite{abdo09}
we can consider
using the spread in arrival times of
photons as a possible technique for detecting spacetime foam.
Now, the spread of arrival times can be
traced to fluctuations in the distance that the photons have travelled from
the distant source to our telescopes.  Hence, according to the 
spacetime foam model parametrized by $\alpha$,  we get
\begin{equation*}
\delta t \sim t^{1 - \alpha} t_P^{\alpha} \sim \delta l / c 
\end{equation*}
{\it for the spread in arrival time of the
photons}, \cite{ng08} independent of energy $E$ (or photon wavelength
$\lambda$). Here $t_P \sim 10^{-44}$ sec is the minuscule Planck time.
Thus the result is that the time-of-flight differences
increase only with the $(1 - \alpha)$-power of the
average overall time of travel $t = l/c$ from the gamma ray bursts to our
detector, {\it leading to a time spread too small to be detectable} 
(except for 
the uninteresting range of $\alpha$ close to 0.)
The new Fermi Gamma-ray Space Telescope results \cite{abdo09}
of $\delta t \lesssim 1$sec for $t \sim 7$ billion years rule out only
spacetime foam models with $\alpha \lesssim 0.3$. The
holographic model predicts an energy independent dispersion of arrival
times $\sim 2.5 \times 10^{-24}$sec.

Thus we see that, while useful in putting a limit on the variation of
the speed of light of a definite sign, this technique is far less useful 
than the measured angular size in constraining the
degree of fuzziness of spacetime in the spacetime
foam models. It is easy to understand why that is the case:
spacetime foam models predict that
the speed of light fluctuates with the fluctuations
taking on $\pm$ sign with equal probability; at one
instant a particular photon is faster than the average
of the other photons, but at the next instant it is
slower.  The end result is that the cumulative effect 
due to spacetime foam on the spread in arrival times
of photons from distant GRBs is very 
small (except for spacetime foam models with small $\alpha$).

\section{Conclusion}

Due to the unity of physics, various physical phenomena 
and structures are inter-related.  In this talk I 
have taken the extreme position of arguing
that spacetime foam is the 
origin of a host of phenomena.  For example,   
the holographic principle finds its roots in spacetime 
foam physics which also sheds light in explaining 
why dark energy/dark matter exists.
Spacetime foam may explain the observed sizes/magnitudes of 
the cosmic energy density, the dynamical cosmological constant and 
the critical galactic acceleration in MoND.  It points to the need for 
cold dark matter with MoND scaling.  Possibly spacetime foam is a
cause of nonlocal gravitational dynamics.  And it has attributes of a 
turbulent fluid.  These are some of the various facets of spacetime foam.
Collectively, these facets provide an interesting picture of (and perhaps
even some indirect evidence for) it. 
For completeness, I should add that an {\it observable}
spread in arrival times for (simultaneously emitted) energetic photons
from gamma-ray bursts is {\it not} among the facets of 
holographic spacetime foam.

\section{Acknowledgments}

I owe much of my understanding of the works
presented in this talk to many people.  They include H.~van Dam, 
M.~Arzano, W.~Christiansen, D.~Floyd,
C.M.~Ho, V.~Jejjala, T.~Kephart, S.~Lloyd, 
D.~Minic, D.~Pavon, E.~Perlman, and C.H.~Tze.  I thank them all.
I thank A. Glindemann for the hospitality extended to 
me at the Headquarters of the European Southern Observatory where most of 
this talk was written while I was on research leave supported by Kenan 
Leave from the University of North Carolina in Fall 2010.  
I also thank L.~Ng for his help in the preparation of this manuscript.
This work was also supported 
in part by the US Department of Energy and by the Bahnson Fund of UNC.

\section{Appendix A: Entropic interpretation of Newton's laws}

In this Appendix we review the recent work of E. Verlinde \cite{verlinde}
in which the canonical Newton's laws are derived
from the point of view of holography.
Using the first law of thermodynamics, Verlinde proposes 
the concept of entropic force
$
F_{entropic} = T \frac{\Delta S}{\Delta x},
$
where $\Delta x$ denotes an infinitesimal spatial displacement of a 
particle with mass $m$ from the heat
bath with temperature $T$. He then
invokes Bekenstein's original arguments
concerning the entropy $S$ of black holes \cite{bekenstein}
by imposing $
\Delta S = 2\pi k_B \frac{mc}{\hbar} \Delta x
$.\,
Using the famous formula for the Unruh temperature,
$
k_B T = \frac{\hbar a}{ 2\pi c},
$\,
associated with a uniformly accelerating (Rindler) observer 
\cite{Untemp},
he obtains 
\begin{equation*}
F_{entropic}= T \nabla_x S= m a,
\end{equation*} 
Newton's second law (with the vectorial form
$
\vec{F} = m \vec{a},
$\,
being dictated by the gradient of the entropy).

Next, Verlinde considers an
imaginary quasi-local (spherical) holographic screen of area $A=4 \pi r^2$ 
with
temperature $T$. Then, he assumes the equipartition of energy $E= 
\frac{1}{2} N k_B T$ with $N$ being
the total number of degrees of freedom (bits) on the screen given by $N = 
Ac^3/(G \hbar)$. Using the Unruh
temperature formula and the fact that $E=M c^2$, he obtains
\begin{equation*}
2 \pi k_B T = G M /r^2
\end{equation*}
and recovers exactly the non-relativistic Newton's law
of gravity, namely $a= G M /r^2$. Note that this is precisely the 
fundamental relation that Milgrom
is proposing to modify so as to fit the galactic rotation curves.

\section{Appendix B: Infinite statistics}

What is infinite statistics?  Succinctly, a Fock realization of infinite
statistics \footnote{More generally, infinite statistics is realized
by a $q$ deformation of the commutation relations
of the oscillators:
$a_k a^{\dagger}_l - q a^{\dagger}_l a_k = \delta_{kl}$ with $q$ between
-1
and 1 (the case $q = \pm 1$ corresponds to bosons or 
fermions).\cite{greenberg}} 
is given by the average of the commutation relations of the 
bosonic and fermionic oscillators
\begin{equation*}
a_k a^{\dagger}_l = \delta_{kl}.
\end{equation*}
Two states
obtained by acting with the $N$ oscillators in different orders are
orthogonal.  It follows that the states may be in any representation
of the permutation group.  The statistical mechanics of particles obeying
infinite statistics can be obtained in a way similar to Boltzmann
statistics, with the crucial difference that the Gibbs
$1/N!$ factor is absent for the former.  Infinite statistics can be
thought of as corresponding to the statistics of identical particles with 
an
infinite number of internal degrees of freedom, which is
equivalent to the statistics of nonidentical particles since they are
distinguishable by their internal states.

As mentioned in the text, a theory of particles
obeying infinite statistics cannot be local \cite{fredenhagen,
greenberg}.  
(That is, the fields associated with infinite statistics are not local,
neither in the sense that their observables commute at spacelike 
separation
nor in the sense that their observables are pointlike functionals of the 
fields.)  
The expression for the number operator 
is both nonlocal and nonpolynomial in the field operators,
and so is the Hamiltonian.  The lack of
locality may make it difficult to formulate a relativistic verion of the
theory; but it appears that a non-relativistic theory can be developed.
Lacking locality also means that the familiar spin-statistics relation is 
no
longer valid for particles obeying infinite statistics; hence
they can have any spin.  Remarkably, the TCP theorem and cluster
decomposition have been shown to hold despite the lack of locality.
\cite{greenberg}


\begin{thebibliography}{99}
\raggedright
\footnotesize

\bibitem{wig57}
E.P. Wigner,
Rev. Mod. Phys. 29 (1957) 255.

\bibitem{sal58}
H. Salecker \& E.P. Wigner,
Phys. Rev. 109 (1958) 571.

\bibitem{ng94}
Y.J. Ng \& H. van Dam,
Mod. Phys. Lett. A9 (1994) 335.

\bibitem{ng95}
Y.J. Ng \& H. van Dam,
Mod. Phys. Lett. A10 (1995) 2801.

\bibitem{Karol}
See also F. Karolyhazy, 
Il Nuovo Cimento A42 (1966) 390; N. Sasakura, Prog. Theor. 
Phys. 102 (1999) 169; A. Frenkel, Found. Phys. 20 (1990) 159.


\bibitem{llo04}
S. Lloyd \& Y.J. Ng,
Scientific American 291, \#5 (2004) 52.

\bibitem{gio04}
V. Giovannetti, S. Lloyd, \& L. Maccone, Science 306 (2004) 1330.

\bibitem{mar98}
N. Margolus \& L.B. Levitin,
Physica (Amsterdam) 120D (1998) 188.

\bibitem{hsu}
Note the qualification ``average''.  This result is consistent
with X. Calmet, M. Graesser, \& S.D. Hsu, 
Phys. Rev. Lett. 93 (2004) 211101.

\bibitem{FordHu}
For related effects of quantum fluctuations
of spacetime geometry, see, e.g.,
L.H. Ford, 
Phys. Rev. D 51 (1995) 1692;
B.L. Hu \& K. Shiokawa, 
Phys. Rev. D 57 (1998) 3474.

\bibitem{AC}
G. Amelino-Camelia, 
Nature 398 (1999) 216.


\bibitem{lie03}
R. Lieu \& L.W. Hillman,
ApJ 585 (2003) L77.

\bibitem{NCvD}
Y.J. Ng, W. Christiansen \& H. van Dam,
Astrophys. J. 591 (2003) L87.

\bibitem{per02}
E.S. Perlman et al.,
AJ 124 (2002) 2401.

\bibitem{chr06}
W. Christiansen, Y.J. Ng \& H. van Dam, Phys. Rev. Lett. 96
(2006) 051301.

\bibitem{cfnp}
W.A. Chritiansen, D. Floyd, Y.J. Ng \& E. Perlman,
arXiv:0912.0535.

\bibitem{GWinterf}
Y.J. Ng, Mod. Phys. Lett. A18 (2003) 1073;
Y.J. Ng, Int. J. Mod. Phys. 11 (2002) 1585; E. Goklu \& C. Lammerzahl,
arXiv:0908.3797.

\bibitem{wheeler}
J.A. Wheeler in {\it Relativity, Groups and Topology}, 
B.S. DeWitt \& C.M. DeWitt, Eds. (Gordon \& Breach, New York,
1963) 315.

\bibitem{unruh}
W.~Unruh,
Phys.\ Rev.\ Lett.\ 46 (1981) 1351;
Phys.\ Rev.\ D 51 (1995) 282.

\bibitem{abh}
M.~Novello, M.~Visser, and G.~Volovik, Eds., {\it Artificial black holes}
(World Scientific, Singapore,2002).

\bibitem{kol}
A.~N.~Kolmogorov,
Dokl.\ Akad.\ Nauk SSSR 30 (1941) 299.

\bibitem{previous}
V.~Jejjala, D.~Minic, Y.~J.~Ng, \& C.~H.~Tze,
Class.\ Quant.\ Grav.\ 25 (2008) 225012;
Mod. Phys. Lett. A25 (2010) 2541;
Int. J. Mod. Phys. D19 (2010) 2311.

\bibitem{wbhts}
G. 't Hooft in {\it Salamfestschrift},
A. Ali et al., Eds.
(World Scientific, Singapore, 1993) 284;
L. Susskind,
J. Math.
Phys. (N.Y.) 36 (1995) 6377;
S.B. Giddings, 
Phys. Rev. D 46 (1992) 1347;
R. Bousso, 
Rev. Mod. Phys. 74 (2002) 825.

\bibitem{bekenstein}
J. D. Bekenstein, Phys. Rev. D7 (1973) 2333.

\bibitem{hawking}
S.W. Hawking, Comm. Math. Phys. 43 (1975) 199.



\bibitem{found}
Y.J. Ng \& H. van Dam, 
Found. Phys. 30 (2000) 795.


\bibitem{Arzano}
M.~Arzano, T.~W.~Kephart, \& Y.~J.~Ng, Phys. Lett. B 649
(2007) 243.

\bibitem{plb}
Y.~J.~Ng, Phys. Lett. B 657 (2007) 10.

\bibitem{DP}
I. Duran \& D. Pavon, arXiv:1012.2986.

\bibitem{GWCS}
C. Gao, F. Wu, X. Chen \& Y.G. Shen, Phys. Rev. D79 (2009) 043511.


\bibitem{sorkin}
R.D. Sorkin, Int. J. Theo. Phys. 36 (1997) 2759.


\bibitem{unimod}
Y.~J.~Ng \& H. van Dam, Phys. Rev. Lett. 65 (1990) 1972; J.J. van der 
Bij, H. van Dam \& Y. J. Ng, Physica A116 (1982) 307;
L. Smolin, arXiv:0904.4841.


\bibitem{diego}
See, e.g., A. Berera, I.G. Moss \& R.O. Ramos, arXiv:0808.1855.

\bibitem{pavon}
W. Zimdahl \& D. Pavon, arXiv:0606555 [astro-ph]; D. Pavon \& B. Wang, 
arXiv:0712.0565.  Also see C. Wetterich, Nucl. Phys. B302 (1988) 668.


\bibitem{verlinde}
E. Verlinde,
arXiv:1001.0785; also see T. Jacobson, Phys. Rev. Lett.
75 (1995) 1260;
T. Padmanabhan, arXiv:0912.3165;
L. Smolin, arXiv:1001.3668.


\bibitem{HMN}
C. M. Ho, D. Minic \& Y. J. Ng, 
Phys. Lett. B693 (2010) 567 [arXiv:1005.3537].

\bibitem{Untemp}
P.C.W. Davies, J. Phys. A8 (1975) 609; 
W.G. Unruh, Phys. Rev. D14 (1976) 870.


\bibitem{deser}
S. Deser \& O. Levin, Class. Quant. Grav. 14 (1997) L163;
T. Jacobson, Class. Quant. Grav. 15 (1998) 251.


\bibitem{dsmond}
See, e.g., M. Milgrom, Astrophys. J. 698 (2009) 1630.


\bibitem{mond}
M. Milgrom, Astrophys. J. 270 (1983) 365, 371, 384;
J. D. Bekenstein, Phys. Rev. D70 (2004) 083509.


\bibitem{interpol}
M. Milgrom, Phys. Lett. A253 (1999) 273.

\bibitem{Mann}
P.D. Mannheim, arXiv:1101.2186;
J.W. Moffat, arXiv:1101.1935.

\bibitem{dark}
For a recent review, see G. Bertone, D. Hooper \& J. Silk,
Phys. Rept. 405 (2005) 279, and references therein.

\bibitem{DHR}
S. Doplicher, R. Haag \& J. Roberts, 
Commun. Math. Phys. 23 (1971) 199;
35 (1974) 49;
A.B. Govorkov, 
Theor. Math. Phys. 54 (1983) 234.

\bibitem{greenberg}
O.W. Greenberg, 
Phys. Rev. Lett. 64 (1990) 705.

\bibitem{CCL}
Also see C. Cao, Y.X. Chen \& J.L. Li, Phys. Rev. D80 (2009) 125019.


\bibitem{minic}
V. Jejjala, M. Kavic \& D.~Minic, 
arXiv:0705.4581.

\bibitem{fredenhagen}
K. Fredenhagen, 
Commun. Math. Phys. 79 (1981) 141.


\bibitem{SG}
S.B. Giddings,
 arXiv:0705.2197;
G.T. Horowitz,
arXiv:0708.3680.


\bibitem{abdo09}
A.A. Abdo et al., Nature 462 (2009) 331.

\bibitem{ng08}
Y.J. Ng, Entropy 10 (2008) 441 [arXiv:0801.2962]


\end{thebibliography}
\end{document}